\documentclass[conference]{IEEEtran}
\IEEEoverridecommandlockouts
\usepackage{cite}
\usepackage{amsmath,amssymb,amsfonts}
\usepackage{algorithmic}
\usepackage{graphicx}
\usepackage{textcomp}
\usepackage{xcolor}
\usepackage{url}
\usepackage{pgf-umlcd}
\usetikzlibrary{backgrounds}
\usetikzlibrary{shapes}
\usetikzlibrary{decorations.pathmorphing}

\pgfdeclarelayer{bg}    
\pgfdeclarelayer{bg2}    

\renewcommand{\umldrawcolor}{black}

\usepackage{rotating}
\usepackage{multirow}
\usepackage{comment}
\usepackage{booktabs}

\usepackage{diagram-tools}

\newcommand*\circlew[1]{\raisebox{.5pt}{\textcircled{\raisebox{-.9pt} {#1}}}}

\usepackage[colorlinks=true,allcolors=black]{hyperref}

\graphicspath{{figures/}}

\usepackage{xspace}
\newcommand*{\eg}{e.g.,\@\xspace}
\newcommand*{\ie}{i.e.\@\xspace}

\newcommand*{\aka}{a.k.a.\xspace}

\newcommand*\xbar[1]{%
  \hbox{%
    \vbox{%
      \hrule height 0.1pt %
      \kern0.4ex
      \hbox{%
        \kern-0.1em%
        \ensuremath{#1}%
        \kern-0.1em%
      }%
    }%
  }%
}

\DeclareMathOperator*{\argmin}{arg\,min}

\usepackage[record,acronym]{glossaries-extra}
\glsxtrsetpopts{hyper=false}
\glsxtrsetpopts{noindex}
\glsdisablehyper
\setabbreviationstyle[acronym]{long-short}
\newacronym[shortplural={DCOPs},longplural={distributed constraint optimization problems}]{dcop}{DCOP}{distributed constraint
optimization problem}
\newacronym[shortplural={DisCSPs},longplural={distributed constraint satisfaction problems}]{dcsp}{DisCSP}{distributed constraint
satisfaction problem}

\newacronym[shortplural={CSPs},longplural={constraint satisfaction problems}]{csp}{CSP}{constraint satisfaction problem}

\newacronym[shortplural={COPs},longplural={constraint optimization problems}]{cop}{COP}{constraint optimization problem}

\newacronym{mape}{MAPE}{monitor, analyze, plan, and execute}
\newacronym{mapek}{MAPE-K}{monitor, analyze, plan, execute, and knowledge}

\newacronym{map}{MAP}{multi-agent planning}
\newacronym{mapc}{MAPC}{multi-agent plan coordination}

\newacronym{ml}{ML}{machine learning}

\newacronym{iot}{IoT}{internet of things}

\newacronym{qos}{QoS}{quality of service}
\newacronym{qoe}{QoE}{quality of experience}

\newacronym[shortplural={SLAs}, longplural={service-level agreements}]{sla}{SLA}{service-level agreement}

\newacronym{p2p}{P2P}{peer-to-peer}

\newacronym{cps}{CPS}{cyber-physical systems}

\newacronym{mec}{MEC}{multi-access edge}

\newglossaryentry{cloud}{name=cloud, description={Cloud computing}}

\newglossaryentry{edge}{name=edge, description={edge computing}}
\newglossaryentry{fog}{name=fog, description={fog computing}}

\newacronym{im}{IM}{infrastructure manager}

\newacronym[shortplural={AMs}, longplural={application managers}]{am}{AM}{application manager}

\newacronym{saas}{SaaS}{software-as-a-service}
\newacronym{iaas}{IaaS}{infrastructure-as-a-service}
\newglossaryentry{sas}{name={self-adaptive system}, description={self-adaptive system}, plural={self-adaptive systems}}
\newacronym{rl}{RL}{reinforcement learning}
\newacronym{marl}{MARL}{multi-agent reinforcement learning}
\newacronym{plancoord}{PC}{plan coordination}
\newacronym{analysiscoord}{AC}{analysis coordination}
\newacronym{mdp}{MDP}{Markov Decision Process}

\newglossaryentry{coordpoint}{name={coordination point}, description={coordination point}, plural={coordination points}}

\newglossaryentry{coordobjective}{name={coordination objective}, description={coordination objective}, plural={coordination objectives}}

\newglossaryentry{subcoordobjective}{name={coordination sub-objective}, description={coordination sub-objective}, plural={coordination sub-objectives}}

\newglossaryentry{coordpeer}{name={coordination peer}, description={coordination peer}, plural={coordination peers}}

\newglossaryentry{directcoordpeer}{name={direct coordination peer}, description={direct coordination peer}, plural={direct coordination peers}}

\newglossaryentry{coordalg}{name={coordination mechanism}, description={coordination algorithm}, plural={coordination algorithms}}

\newglossaryentry{coordpointA}{name={analysis coordination point}, description={analysis coordination point}, plural={analysis coordination points}}
\newglossaryentry{coordpointAPC}{name={trigger coordination point}, description={trigger coordination point}, plural={trigger coordination points}}

\newglossaryentry{coordpointPstate}{name={state coordination point}, description={state coordination point}, plural={state coordination points}}
\newglossaryentry{coordpointPaction}{name={action coordination point}, description={action coordination point}, plural={action coordination points}}

\newglossaryentry{coordconstA}{name={analysis consistency objective}, description={}}
\newglossaryentry{coordconstApref}{name={adaptation preference objective}, description={}}

\newglossaryentry{coordconstP}{name={state consistency objective}, description={}}
\newglossaryentry{coordconstPpref}{name={state preference objective}, description={}}
\newglossaryentry{coordconstPA}{name={action consistency objective}, description={}}
\newglossaryentry{coordconstPApref}{name={action preference objective}, description={}}
\newglossaryentry{coordconstPPA}{name={state-action consistency objective}, description={}}

\newglossaryentry{unthreaded}{name={unthreaded planning and coordination}, description={}}
\newglossaryentry{interleaved}{name={interleaved planning and coordination}, description={}}
\newacronym[shortplural={VMs}, longplural={virtual machines}]{vm}{VM}{virtual machine}

\newacronym{admm}{ADMM}{alternating direction method of multipliers}
\newacronym{mip}{MIP}{mixed integer programming}
\newacronym{des}{DES}{discrete-event systems}
\newacronym{eca}{ECA}{event-condition-action}
\newacronym{ltl}{LTL}{linear temporal logic}

\newacronym[shortplural={MS}, longplural={managing systems}]{ms}{MS}{managing system}
\newacronym[shortplural={MEs}, longplural={managed elements}]{me}{ME}{managed element}
\newacronym[shortplural={ALs}, longplural={adaptation logics}]{al}{AL}{adaptation logic}
\newacronym{bsm}{BSM}{behavior specification manager}
\newacronym[shortplural={PSMs}, longplural={planning strategy managers}]{psm}{PSM}{planning strategy manager}
\newglossaryentry{aps}{name={adaptation planning strategy}, description={adaptation planning strategy}, plural={adaptation planning strategies}}

\newcommand{\citeauthor}[1]{{\color{red} authorname}\xspace}

\usepackage[inline]{enumitem}

\def\BibTeX{{\rm B\kern-.05em{\sc i\kern-.025em b}\kern-.08em
    T\kern-.1667em\lower.7ex\hbox{E}\kern-.125emX}}
\begin{document}

\title{Towards the decentralized coordination of multiple self-adaptive systems\\\thanks{
Research leading to these results  received funding from the EU's Horizon 2020 and Horizon Europe R\&I programmes under grant agreements 871525 (FogProtect) and 101070455 (DynaBIC).
}}

\author{\IEEEauthorblockN{Paul-Andrei Dragan, Andreas Metzger, Klaus Pohl}
\IEEEauthorblockA{\textit{paluno (The Ruhr Institute for Software Technology)} \\
\textit{University of Duisburg-Essen}, Essen, Germany \\
paul-andrei.dragan@paluno.uni-due.de, andreas.metzger@paluno.uni-due.de, klaus.pohl@paluno.uni-due.de}
}

\newcommand{\refsec}[1]{Section {\ref{#1}}\xspace}
\newcommand{\reffig}[1]{Figure {\ref{#1}}\xspace}
\newcommand{\reftab}[1]{Table {\ref{#1}}\xspace}
\newcommand{\refeq}[1]{{\eqref{#1}}\xspace}

\newcommand{\am}[1]{{\color{red}\textbf{AM:}~{#1}}}
\newcommand{\pd}[1]{{\color{blue}\textbf{PD:}~{#1}}}
\newcommand{\todo}{{\color{red}\textbf{TODO}}\xspace}
\newcommand{\TODO}[1]{{\color{red}[\textbf{TODO} #1]}\xspace}

\newcommand{\dcop}{\gls{dcop}\xspace}
\newcommand{\dcops}{\glspl{dcop}\xspace}
\newcommand{\csp}{\gls{csp}\xspace}
\newcommand{\csps}{\glspl{csp}\xspace}
\newcommand{\cop}{\gls{cop}\xspace}
\newcommand{\cops}{\glspl{cop}\xspace}
\newcommand{\mape}{\gls{mapek}\xspace}
\newcommand{\runtime}{runtime\xspace}
\newcommand{\designtime}{design-time\xspace}
\newcommand{\ml}{\gls{ml}\xspace}
\newcommand{\iot}{\gls{iot}\xspace}
\newcommand{\cloud}{\gls{cloud}\xspace}
\newcommand{\dsas}{decentralized self-adaptive system\xspace}
\newcommand{\dsases}{decentralized self-adaptive systems\xspace}
\newcommand{\sas}{self-adaptive system\xspace}
\newcommand{\sases}{self-adaptive systems\xspace}
\newcommand{\im}{\gls{im}\xspace}
\newcommand{\appm}{\gls{am}\xspace}
\newcommand{\appms}{\glspl{am}\xspace}
\newcommand{\sota}{state-of-the-art\xspace}
\newcommand{\masterslave}{Master/Slave\xspace}
\newcommand{\mapef}[1]{\MakeUppercase{#1}\xspace}
\newcommand{\coordmon}{coordinated monitoring\xspace}
\newcommand{\plancoord}{\gls{plancoord}\xspace}
\newcommand{\analysiscoord}{analysis coordination\xspace}
\newcommand{\astar}{A*\xspace}

\newcommand{\CoordDCOP}{CoADAPT\xspace}
\newcommand{\ACDCOP}{\CoordDCOP-AC\xspace}
\newcommand{\PCDCOP}{\gls{plancoord}\xspace}

\newcommand{\forwardp}{\textit{aggregation} problem\xspace}
\newcommand{\backwardp}{\textit{distribution} problem\xspace}
\newcommand{\coordman}{coordination management\xspace}
\newcommand{\preexecp}{\textit{pre-execution} problem\xspace}
\newcommand{\postexecp}{\textit{post-failure} problem\xspace}
\newcommand{\failurep}{\textit{failure handling} problem\xspace}
\newcommand{\conarch}{coordination conceptual architecture\xspace}
\newcommand{\layerMS}{\textit{managed system} layer\xspace}
\newcommand{\layerMAPE}{\textit{managing system} layer\xspace}
\newcommand{\layerDecen}{\textit{decentralization} layer\xspace}
\newcommand{\layerDCOP}{\textit{coordination} layer\xspace}
\newcommand{\layerCoordMan}{\textit{coordination management} layer\xspace}

\newcommand*{\ICeff}{(B-1)\xspace}
\newcommand*{\ICperf}{(B-2)\xspace}
\newcommand*{\ACeff}{(A-1)\xspace}
\newcommand*{\ACperf}{(A-2)\xspace}

\newcommand*{\SVone}{SV1\xspace}
\newcommand*{\SVtwo}{SV2\xspace}

\newcommand{\pydcop}{pyDcop\xspace}

\maketitle

\begin{abstract}
When multiple self-adaptive systems share the same environment and
have common goals, they may coordinate their adaptations at
\runtime to avoid conflicts and to satisfy their
goals. There are two approaches to coordination. (1) Logically
centralized, where a supervisor has complete control over the
individual self-adaptive systems. Such approach is infeasible when the
systems have different owners or administrative domains. (2) Logically
decentralized, where coordination is achieved through direct
interactions. Because the individual systems have control over the
information they share, decentralized coordination accommodates
multiple administrative domains. However, existing techniques do not
account simultaneously for both local concerns, \eg preferences, and
shared concerns, \eg conflicts, which may lead to goals not being
achieved as expected. Our idea to address this shortcoming is to
express both types of concerns within the same constraint optimization
problem. We propose \CoordDCOP, a decentralized coordination technique
introducing two types of constraints: preference constraints,
expressing local concerns, and consistency constraints, expressing
shared concerns. At \runtime, the problem is solved in a decentralized
way using distributed constraint optimization algorithms implemented
by each self-adaptive system. As a first step in realizing \CoordDCOP,
we focus in this work on the coordination of adaptation planning
strategies, traditionally addressed only with centralized
techniques. We show the feasibility of \CoordDCOP in an exemplar
from \gls{cloud} computing and analyze experimentally its scalability.

\end{abstract}



\newcommand{\limitation}[1]{\textit{Limitation #1}\xspace}
\newcommand{\limitationOne}{\limitation{1}}
\newcommand{\limitationTwo}{\limitation{2}}
\newcommand{\limitationThree}{\limitation{3}}

\section{Introduction} \label{sec:NewIntro}




A \gls{sas} modifies its structure or behavior during operation to
continuously meet its goals and requirements
\cite{Weyns_2020}. Conceptually, a \gls{sas} can be structured into
two main parts: a \gls{me}, realizing the domain logic, and a
\gls{ms}, responsible for the adaptation
concerns\cite{Weyns_2013,Weyns_2020}.

When multiple self-adaptive systems share an environment and have
common goals, they may coordinate their adaptations at \runtime to
avoid conflicting adaptations and to satisfy their common goals
\cite{Weyns_2013}. For example, in \gls{cloud} computing, multiple
self-adaptive applications may be deployed on a shared self-adaptive
infrastructure; uncoordinated adaptations, \eg uncoordinated
elasticity, may lead to reduced performance \cite{Chen_Bahsoon2_2017}.

Coordination refers to joint \runtime activities carried out by the
\glspl{sas}, at the level of their managing systems, for aligning on
adaptation. Such activities can include joint goal management
\cite{Wohlrab_2022}, adaptation planning
\cite{Calinescu_2015,Tsigkanos_2019,metzger2016coordinated,Baresi_2019,Gerostathopoulos_2019,DAngelo_2020},
or adaptation enactment \cite{Matusek_2022}, among others.

There are two main approaches to coordination. One approach is
logically centralized, where a supervisor has complete knowledge and
control of the individual self-adaptive systems, \eg an
\textit{application manager} coordinating the elasticity of
microservices \cite{Rossi_2020}. This approach may be infeasible,
particularly in cases where the systems have different owners or
administrative domains, \eg large-scale compositions of web services
\cite{DAngelo_2020}, cloud federations \cite{Carlini_2016}, or smart
grids \cite{Guo_Miao_2013}.

The other approach is logically decentralized, where coordination is
achieved via direct communication between the self-adaptive systems,
\eg by exchanging information on their monitored local state
\cite{DAngelo_2020,Azlan_2015}. Because the individual systems have
more control over their adaptations and information they share,
decentralized coordination accommodates multiple administrative
domains and may improve privacy. However, existing techniques do not
properly handle simultaneously both local concerns, \eg the preference
for a particular adaptation in the case one \gls{sas}, and shared
concerns, \eg conflicts between adaptations pertaining to multiple
\glspl{sas}.

This limitation is due to at least one of the following
reasons. First, existing techniques may not explicitly differentiate
between the two types of concerns, leading to the sharing of private
information \cite{DAngelo_2020,Azlan_2015,Calinescu_2015}. Second,
they may make unrealistic assumptions about coordination, \eg there
are no conflicts between adaptations
\cite{metzger2016coordinated}. Third, these techniques may rely on
local decision-making only (in contrast to joint decision-making)
which may lead to common goals not being achieved as expected
\cite{Calinescu_2015,DAngelo_2020}

Our idea to address these shortcomings, is to express both local and
shared concerns within the same constraint optimization problem. Thus,
we propose \CoordDCOP, a decentralized coordination technique for
self-adaptive systems, introducing two types of constraints:
preference constraints, expressing local concerns, and consistency
constraints, expressing shared concerns. At \runtime, the problem is
solved in a decentralized way using distributed algorithms implemented
by each \gls{sas}. With \CoordDCOP, the \glspl{sas} engage in joint
decision-making, conflicts or other shared concerns are explicitly
accounted for, and information about local concerns remains private
during coordination.

As a suitable base for \CoordDCOP we have identified the \gls{dcop}
formalism. \Glspl{dcop} \cite{Modi_2005,petcu2005dpop,Fioretto_2018}
are a formal problem setting from the field of multi-agent systems,
where agents coordinate the values assigned to the variables they
control so as to optimize a global objective. \Glspl{dcop} were
studied for many applications, \eg smart grids
\cite{fioretto2017multiagent}, logistics
\cite{leaute2011coordinating}, or cloud computing \cite{Jin_2011}.

The benefits of \gls{dcop} for coordinating \glspl{sas} are
threefold. First, \glspl{dcop} can naturally express decentralized or
distributed coordination via the construct of \textit{agents}. Second,
\glspl{dcop} can express both local and shared concerns using
constraints. Third, a number of algorithms exist to solve \glspl{dcop}
that keep certain constraints private to individual agents
\cite{Fioretto_2018}. Moreover, these algorithms offer different
\runtime characteristics, \eg space and time complexities,
accommodating various system design requirements, \eg different
numbers of coordinating \glspl{sas} \cite{Fioretto_2018}.

As a first step in realizing \CoordDCOP, we focus in this work on
coordinating the assignment of adaptation planning strategies
\cite{Lesch_Hadry_2022}, traditionally addressed only with centralized
techniques \cite{Wang_Ying_Harmonizing_2007,
  Khakpour_2010,Glazier_2019, Glazier_2020}. In contrast to
coordination at the level of adaptation planning
\cite{Calinescu_2015,Tsigkanos_2019,metzger2016coordinated,Baresi_2019,Gerostathopoulos_2019,DAngelo_2020},
coordinating at the higher level of \glspl{aps} has the important
benefit of better supporting heterogeneity \cite{Glazier_2019,
  Glazier_2020}, \eg adaptation logics could be realized by means of
different techniques or operate at different timescales. Coordination
at such a higher level has various applications, \eg assigning
elasticity policies to individual \gls{cloud} applications for better
resource utilization \cite{Rossi_2020} or orchestrating security
policies \cite{Molina_2020}.

To summarize, our contribution in this paper consists of:
\begin{itemize}
\item We propose an architecture for the decentralized coordination of
  \glspl{aps}.
\item We propose a formalization for the coordination of strategies
  taking into account both local and shared concerns.
\item We describe how to map such formalization to a \gls{dcop}.
\item We describe a process for selecting an algorithm for realizing
  coordination at \runtime.
\item We show the feasibility of \CoordDCOP in the Simdex
  \cite{Krulis_2022} exemplar and analyze experimentally its
  scalability.
\end{itemize}

The rest of the paper is structured as follows:
\refsec{sec:Background} gives a brief background on \glspl{dcop} and
on the task of coordinating \glspl{aps}. \refsec{sec:MotivationExmpl}
describes our running example. \refsec{sec:Contribution} describes
decentralized coordination with \CoordDCOP. \refsec{sec:Experimental}
is dedicated to an experimental evaluation. Finally,
\refsec{sec:Related_Work} gives an overview of related work.
\refsec{sec:Conclusion} summarizes our work and discusses potential
future directions.

\section{Background} \label{sec:Background}

\subsection{Distributed constraint optimization problem} \label{sec:BackgroundDCOP}


A \glsxtrfull{dcop} \cite{Modi_2005,Fioretto_2018} is formally defined
as a tuple $\langle A, X, D, F, \pi, \mu \rangle$, where:



\begin{itemize}
  \item $A = \{a_1, \dots, a_n\}$ is the set of \emph{agents}.
  \item $X = \{x_1, \dots, x_n\}$ is the set of \emph{variables}.
  \item $D = \{D_1, \dots, D_n\}$ is a set of finite variable
    \emph{domains}, with $D_i = \{\delta_1, \dots, \delta_p\}$ being the domain of $x_i$.
  \item $F = \{f_1, \dots, f_m\}$ is a set of \textit{constraint
    functions} (a.k.a.\ \textit{cost functions} or
    \textit{constraints}), $f_i: D_{l} \times \cdots \times D_{k} \to
    \mathbb{R^+} \cup \{\infty\}$.  Each $f_i$ has a \emph{scope} $Z_i
    \subseteq X$.  A constraint $f_i$ maps a partial assignment of
    variables $\delta_{Z_i} \in D_{l} \times \cdots \times D_{k}$ from
    its scope $Z_i$ to a real-valued \textit{cost}. The cardinality of
    $Z_i$ is called the \textit{arity} of $f_i$.


  \item $\pi : X \to A$ maps variables to agents; $\pi(x_i) = a_i$
    means that $a_i$ controls how the values from $D_i$ are assigned
    to $x_i$, \ie $a_i$ {controls} $x_i$. For simplicity \dcop
    algorithms assume a one-to-one mapping between agents and
    variables. We do the same in this paper.


\item $\mu(\delta)$ is the \textit{objective function}, where $\delta
  \in D_1 \times \cdots \times D_n$ is a complete assignment. It is
  usually assumed that the objective function is just the sum of
  costs, \ie $\mu (\delta) = \sum_{f_i \in F} f_i(\delta_{Z_i})$ -- a
  key assumption for many algorithms.
\end{itemize}

Solving the \dcop means finding a $\delta^{\star}$ such that:

\vspace{-4mm}
\begin{equation} \label{eq:DCOPmuObjective}
  \delta^{\star} = \argmin_{\delta \in D_1 \times \cdots \times
    D_n} \mu(\delta) = \argmin_{\delta \in D_1 \times \cdots \times
    D_n} \sum_{f_i \in F} f_i(\delta_{Z_i}) \,.
\end{equation}



\glspl{dcop} are solved in a distributed way: agents exchange messages
with neighbors\footnotemark{}, following certain \textit{algorithms}
so as to discover the optimal assignment $\delta^{\star}$. Many
algorithms exist to solve \dcops; a comprehensive survey is provided
in \cite{Fioretto_2018}.

\footnotetext{Two agents $a_1$ and $a_2$ controlling variables $x_1$
  and $x_2$, respectively, are \textit{neighbors} if $\exists f \in F$ with
  scope $Z$ having the property that $x_1, x_2 \in Z$.}

\subsection{Coordination of adaptation planning strategies} \label{sec:BackAPS}

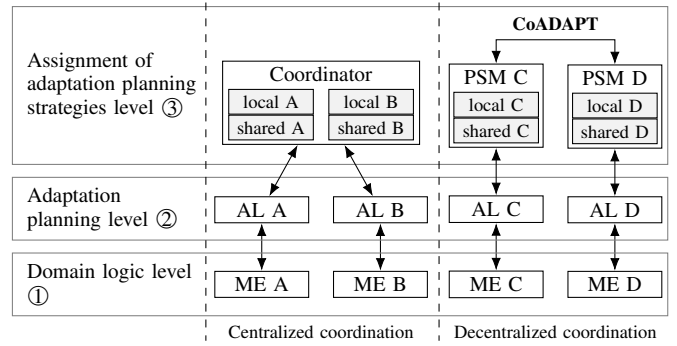
\begin{figure}[t]
  \centering
  \newcommand{\drawarchmapecontrol}[2]{
  \draw [mminter] (#1.south) -- (#2.north);
}

\begin{tikzpicture}[
  scale=1.0,
  transform shape,
  shift=(current page.center),
  node distance=4ex and 2ex,
  font=\rmfamily\footnotesize,
  managedsysarch/.style={
    managedsys,
    text width=3.1em,
    align=center,
    node distance=4ex and 2ex,
    minimum height=1ex,
    inner sep=0.5ex,
    fill=white,
    font=\rmfamily\footnotesize
  },
  managingsysarch/.style={
    managingsys,
    sharp corners,
    align=center,
    minimum height=1ex,
    text width=3.1em,
    inner sep=0.5ex,
    node distance=4ex and 2ex,
    fill=white,
    font=\rmfamily\footnotesize
  },
  labelstyle/.style={
    text width=6.5em,
    align=left
  },
  mminter/.style={
    {Latex[]}-{Latex[]}
  }
  ]


  \node[labelstyle, anchor=west] (SysLayerText) at (0, 0) {Domain logic level \circlew{1}};


  \node[managedsysarch, node distance=2ex, right=0.5ex of SysLayerText] (Managed1)  {ME A};
  \node[managedsysarch, right=of Managed1] (Managed2) {ME B};
  \node[managedsysarch, right=1.75ex of Managed2] (Managed3) {ME C};
  \node[managedsysarch, right=of Managed3] (Managed4) {ME D};

  \node[managingsysarch, above=of Managed1] (MAPE1) {AL A};
  \node[managingsysarch, above=of Managed2] (MAPE2) {AL B};
  \node[managingsysarch, above=of Managed3] (MAPE3) {AL C};
  \node[managingsysarch, above=of Managed4] (MAPE4) {AL D};

  \node[managingsysarch, above=of MAPE3, minimum height=7ex] (PSM3) {};
  \node[font=\footnotesize, above=-2.5ex of PSM3] {PSM C};
  \matrix [matrix of nodes, below=-5.5ex of PSM3, nodes={draw, fill=black!5,
      font=\scriptsize, inner sep=0.5ex, text width=2.75em,
      align=center}] {
    | {local C}; | \\
    | {shared C}; | \\
  };

  \node[managingsysarch, above=of MAPE4, minimum height=7ex] (PSM4) {};
  \node[font=\footnotesize, above=-2.5ex of PSM4] {PSM D};
  \matrix [matrix of nodes, below=-5.5ex of PSM4, nodes={draw, fill=black!5,
      font=\scriptsize, inner sep=0.5ex, text width=2.75em,
      align=center}] {
    | {local D}; | \\
    | {shared D}; | \\
  };

  \drawarchmapecontrol{MAPE1}{Managed1}
  \drawarchmapecontrol{MAPE2}{Managed2}
  \drawarchmapecontrol{MAPE3}{Managed3}
  \drawarchmapecontrol{MAPE4}{Managed4}

  \node[managingsysarch, draw=none, fill=none, above=of MAPE1] (Dummy1) {};
  \node[managingsysarch, draw=none, fill=none, above=of MAPE3] (Dummy3) {};

  \draw[mminter] ($ (PSM3.north) + (0ex, 0) $) -- ++(0, 2ex) -| node[near start, above, font=\scriptsize] {\textbf{\CoordDCOP}} ($ (PSM4.north) + (0ex, 0) $);


  \path let
  \p1=(SysLayerText.west),
  \p2=(Managed4.east),
  \p3=(Managed1.south),
  \p4=(Managed1.north)
  in node [
  draw=black!50,
  minimum height=\y4-\y3+3ex,
  minimum width=\x2-\x1-\pgflinewidth+1ex,
  ] (SysLayer) at ($ (\p1)!0.5!(\p2) $) {};

  \path let
  \p1=(SysLayerText.west),
  \p2=(Managed4.east),
  \p3=(Managed1.south),
  \p4=(Managed1.north),
  \p5=(MAPE1.east),
  \p6=(MAPE4.west),
  in node [
  draw=black!50,
  minimum height=\y4-\y3+3ex,
  minimum width=\x2-\x1-\pgflinewidth+1ex,
  above=1ex of SysLayer,
  ] (MAPELayer) {};

  \path let
  \p1=(SysLayerText.west),
  \p2=(Managed4.east),
  \p3=(Managed1.south),
  \p4=(Managed1.north),
  \p5=(Dummy1.east),
  \p6=(Dummy3.west),
  in node [
  draw=black!50,
  minimum height=\y4-\y3+11ex,
  minimum width=\x2-\x1-\pgflinewidth+1ex,
  above=1ex of MAPELayer,
  ] (DecenLayer) {};


  \path let
  \p1=(MAPE1.west),
  \p2=(MAPE2.east),
  \p3=($ (MAPE1.west)!0.5!(MAPE2.east) + (0, 9ex)$)
  in
  node[
    managingsysarch, minimum height=7ex, text width=7em,
  ] (MAPECoord) at (\x3, \y3) {};

  \node[above=-2.5ex of MAPECoord, font=\footnotesize] {Coordinator};

  \matrix [matrix of nodes, below left=-5.5ex and -8.5ex of MAPECoord, nodes={draw, fill=black!5,
      font=\scriptsize, inner sep=0.5ex, text width=2.75em,
      align=center}] {
    | {local A}; | \\
    | {shared A}; | \\
  };

  \matrix [matrix of nodes, below right=-5.5ex and -8.5ex of MAPECoord, nodes={draw, fill=black!5,
      font=\scriptsize, inner sep=0.5ex, text width=2.75em,
      align=center}] {
    | {local B}; | \\
    | {shared B}; | \\
  };

  \path let
  \p1=(Managed1.west),
  \p2=(Managed2.east),
  \p3=($ (Managed1.west)!0.5!(Managed2.east) + (0, -4ex)$)
  in
  node[
    minimum height=3ex, text width=7em, align=center, font=\scriptsize
  ] at (\x3, \y3) {Centralized coordination};

  \path let
  \p1=(Managed3.west),
  \p2=(Managed4.east),
  \p3=($ (Managed3.west)!0.5!(Managed4.east) + (0, -4ex)$)
  in
  node[
    minimum height=3ex, text width=8em, align=center, font=\scriptsize
  ] at (\x3, \y3) {Decentralized coordination};


  \path let
  \p1=(SysLayerText.center),
  \p2=(MAPELayer.north),
  \p3=(MAPELayer.south),
  \p4=($ (MAPELayer.north)!0.5!(MAPELayer.south) $)
  in node [labelstyle] (MAPELayerText) at (\p1 |- \p4 )  {Adaptation planning level \circlew{2}};


  \path let
  \p1=(MAPELayerText.center),
  \p2=(DecenLayer.north),
  \p3=(DecenLayer.south),
  \p4=($ (DecenLayer.north)!0.5!(DecenLayer.south) $)
  in node [labelstyle] (DecenLayerText) at (\p1 |- \p4 )  {Assignment of adaptation planning strategies level \circlew{3}};



  \draw[mminter] (MAPE1) to (MAPECoord);
  \draw[mminter] (MAPE2) to (MAPECoord);
  \draw[mminter] (PSM3) to (MAPE3);
  \draw[mminter] (PSM4) to (MAPE4);

  \draw[dashed] (5.6, 3.75) to (5.6, -0.75);
  \draw[dashed] (2.5, 3.75) to (2.5, -0.75);

\end{tikzpicture}
  \vspace{-0.5cm}
  \caption{Coordination of \glspl{aps} at the conceptual level (ME =
    managed element, AL = adaptation logic, PSM = planning strategy
    manager).}
  \label{fig:ConceptualArch2}
  \vspace{-5mm}
\end{figure}

For a \gls{sas}, an \gls{aps} is a specification of how adaptation
planning, \ie the creation of sequences of adaptations (\aka
adaptation plans), should take place at the level of the adaptation
logic \cite{Gerostathopoulos_Bures_2017,Lesch_Hadry_2022}. Such
strategy could specify, for example, what resources are made available
to a \gls{sas} \cite{Glazier_2019, Glazier_2020}, what planning
algorithm should be employed \cite{Lesch_Hadry_2022}, or which
rule-based adaptation policy should be followed
\cite{Wang_Ying_Harmonizing_2007,Khakpour_2010}.

Conceptually, operations at the level of \glspl{aps}, \eg the
assignment of strategies, are carried out at a level above adaptation
planning \cite{Lesch_Hadry_2022}. The result of coordination is an
assignment of \glspl{aps} to each of the \glspl{sas}
\cite{Glazier_2019, Glazier_2020}.

\reffig{fig:ConceptualArch2} shows the three conceptual levels
\cite{Lesch_Hadry_2022}. Level \circlew{1} is that of the domain or
application logic, including the \glsxtrlongpl{me} of all the
\glspl{sas}. Level \circlew{2} is that of adaptation planning,
realized by the adaptation logics. Finally, level \circlew{3} covers
the assignment of \glspl{aps} to the adaptation logics. Existing works
\cite{Wang_Ying_Harmonizing_2007,Khakpour_2010, Glazier_2019,
  Glazier_2020} realize level \circlew{3} by means of a central
coordinator. This coordinator assumes complete knowledge of the local
and shared concerns of individual \glspl{sas} and has complete control
over the assignment of \glspl{aps}. We introduce in
\refsec{sec:Contribution} a decentralized alternative that does not
require such assumptions.

\section{Running example} \label{sec:MotivationExmpl}

To illustrate \CoordDCOP, we use a \gls{cloud} system as a running
example. Consider two web services part of the same video streaming
website: \SVone provides video content and \SVtwo provides promotional
materials, \ie ads. The two software systems implementing the services
share an infrastructure, \ie they are co-located.  The services are
affected by uncertainties: the maximum capabilities of the
infrastructure, cost of resources, user demand, per-click ad income,
``tolerance'' of users towards low-quality videos, ``willingness'' of
users to click on ads etc. Therefore, the pre-partitioning of
resources at \designtime is not feasible and self-adaptation is
necessary.

\SVone adapts itself according to two \glspl{aps}: \ACeff, which
adjusts video quality within a given amount of available resources,
and \ACperf, which adjusts the amount of resources to ensure an
acceptable video quality. Similarly, \SVtwo has: \ICeff, which adjusts
ad quality, \eg images instead of videos, and \ICperf, which acquires
resources. The common goal of the two services is to maximize revenue.


During low demand the two services prefer \ACperf and \ICperf,
respectively, to maximize user satisfaction and ad clicks. However,
during high demand, there are not enough resources available in the
infrastructure to serve both high-quality videos and high-quality ads.
If there is no coordination during high demand, the two \glspl{sas}
cannot jointly maximize revenue, so they pick their strategies as they
believe appropriate based on local state, \ie (demand, resources,
satisfaction) for \SVone and (demand, resources, clicks) for \SVtwo.
For example, \SVtwo tries to optimize ad clicks by acquiring more
resources \ICperf, while \SVone becomes resources starved so it starts
reducing video quality \ACeff. This could lead to poor user
satisfaction and loss of revenue due to users leaving the system early
and not clicking on ads.

\newcommand*{\iprov}{A}
\newcommand*{\aprov}{B}

\newcommand*\nth[1]{$#1\text{th}$}

\section{Decentralized coordination of adaptation planning strategies with \CoordDCOP} \label{sec:Contribution}

\subsection{Overview} \label{sec:CoordinationProcedure}


To coordinate the assignment of \glspl{aps} between multiple
\glspl{sas} we propose a hierarchical architecture for each of the
coordinating \glspl{sas}, following the three conceptual levels
discussed in \refsec{sec:BackAPS}. Such an architecture enables a
clear separation of concerns within a \gls{sas}
\cite{Braberman_2017,Weyns_2013} and allows individual layers to
operate at different levels of abstraction and at possibly different
timescales \cite{Weyns_2013}.

\begin{figure}[t]
  \centering
  \input{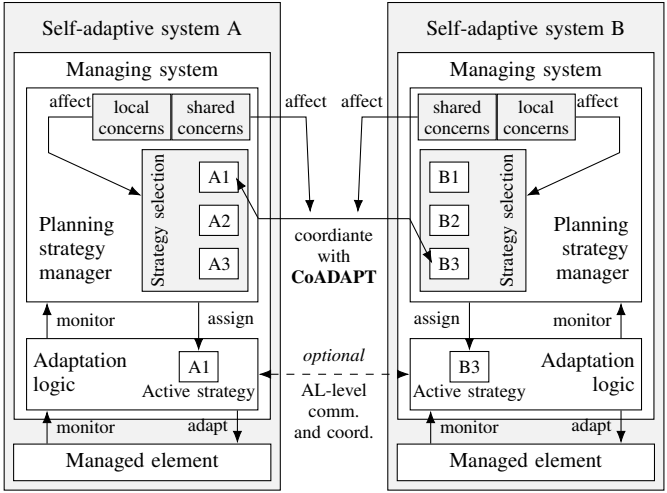}
  \caption{Architecture for coordinating the assignment of \glspl{aps}
    with \CoordDCOP for a particular instance. Arrows show the flow of
    information or of commands.}
  \label{fig:ConceptualArch}
  \vspace*{-3.2mm}
\end{figure}

\reffig{fig:ConceptualArch} depicts our architecture in the case of a
concrete instance of coordination among two \glspl{sas}. The
architecture consists of three main components:
\begin{itemize}
  \item The \glsxtrfull{me} realizes the domain logic of the \gls{sas}
    and is part of the conceptual level \circlew{1}
    (\reffig{fig:ConceptualArch2}).
  \item The \gls{al} responsible for monitoring and adapting
    the \gls{me}. The \gls{al} pertains to a particular the
    \glsxtrfull{ms} and carries out adaptation planning according to
    an active \gls{aps}. The \gls{al} is in conceptual level
    \circlew{2}.
  \item The \gls{psm} is responsible for monitoring the \gls{al} and
    the \gls{me} for selecting an appropriate
    \gls{aps}, and for assigning it to the \gls{al}. The \glspl{psm}
    are part of level \circlew{3} and implement the \CoordDCOP
    \runtime coordination logic.
\end{itemize}


Part of the \gls{psm} are the local concerns, \eg preferences for a
certain \gls{aps}, and shared concerns, \eg known conflicts between
certain \glspl{aps} of different \glspl{sas}. Both the local and
shared concerns influence the strategy assignment process. However,
during coordination the local concerns remain private to a particular
system, while the shared concerns are visible to the systems affected
by those concerns, \ie there is communication only between systems
sharing concerns.

Our architecture is extended to an arbitrary number of \glspl{sas} as
follows: 1) each system will have its own set of local concerns, 2)
each system will have a set of shared concerns for each coordinating
set\footnotemark{}, and 3) direct communication will happen only
within each coordinating set.

\footnotetext{A \textit{coordinating set} includes all \glspl{sas}
  sharing a concern.}


In this paper, we assume that the following exist when realizing
the architecture from \reffig{fig:ConceptualArch}:

\begin{itemize}
  \item A trusted system integrator responsible for overseeing the
    design and specification of coordination.
  \item A fixed set of \glspl{sas} and each system has a fixed set of
    possible strategies.
  \item Perfect communication during coordination.
  \item Cooperation among the systems in achieving common goals (in
    spite of keeping local concerns private).
  \item (Optional) Further communication between \glspl{al} to
    exchange information relevant for coordination, if necessary
    (dashed arrow in \reffig{fig:ConceptualArch})
    \cite{Siqueira_2023}.
  \item Mechanisms at the level of each \gls{psm} to reflect the
    information from the \gls{al} in local and shared concerns, \ie an
    \textit{update function}.

\end{itemize}

Having a trusted system integrator at \designtime is necessary in
order to ensure that coordination is correctly specified. At
\designtime, this system integrator will:

\begin{enumerate}
  \item Formalize the coordination of \glspl{aps} using the framework
    we introduce in \refsec{sec:CoordFormalization},
  \item Map the formalization to a \dcop by following the steps from
    \refsec{sec:MappingToDCOP},
  \item Decide on an appropriate decentralized algorithm for solving
    the \gls{dcop} using the reasoning from
    \refsec{sec:SelectAlgorithm},
  \item Define the update function to update local and shared concerns
    and reflect them in the \dcop.
\end{enumerate}


At \runtime, periodically or triggered by some events, \eg upon major
changes, the following will happen without the intervention of the
system integrator:

\begin{enumerate}
\item The update function reflects the current state and goals in the
  coordination \dcop,
\item The \glspl{psm} follow the algorithm to solve the \dcop and
  obtain the strategy assignments for each \glspl{al},
\item The \glspl{al} activate the \glspl{aps}.
\end{enumerate}

\subsection{Formalizing coordination} \label{sec:CoordFormalization}

\subsubsection{Overview}

At \runtime, the \glspl{sas} select their \glspl{aps} so as to best
meet common goals, while satisfying local and shared concerns.
Therefore, we can formalize the coordinated assignment of \glspl{aps}
as an optimization problem. Given $n$ \glspl{sas} with the
architecture from \reffig{fig:ConceptualArch}, coordination is
formalized by the tuple $\langle S, C, O, K, \Sigma, \rho \rangle$:


\begin{itemize}
\item $S = S_1 \times \cdots \times S_n$, where $S_i$ is
  the set of possible states of the \nth{i} \gls{me} and of its
  environment, as monitored by the \nth{i} \gls{al},
\item $C = C_1 \times \cdots \times C_n$, where $C_i$ is
  the set of possible \glspl{aps} of the \nth{i} \gls{al},
\item $O = O_L \times O_S$, where $O_S$ is a set of \textit{shared
  concerns} and $O_L = O_1 \times \cdots \times O_n$ are \textit{local
  concerns},
\item $K = K_L \times K_S$, $K_S$ is the set of shared knowledge and
  $K_L = K_1 \times \cdots \times K_n$ are local knowledge,
\item $\Sigma = S \times C \times O \times K$ is the set of complete
  states of the system composed of the $n$ \glspl{sas},
\item $\rho: \Sigma \times C \to \mathbb{R^+} \cup \{\infty\}$, where
  $\rho(\sigma, c')$ gives the cost of the new strategies $c' = (c'_1,
  c'_2, \dots, c'_n) \in C$, given $\sigma \in \Sigma$, the current
  state of the system.
\end{itemize}

Note that in this paper, \textit{knowledge} means models, \ie
functions $k_i: O_i \times S_i \times C_i \to \mathbb{R}^+ \cup
\{\infty\}$, $k_i \in K_i$ and $k_{Z}: O_{S} \times S_{Z} \times C_{Z}
\to \mathbb{R}^+ \cup \{\infty\}$, $k_{Z} \in K_S$, expressing the
\textit{contribution} of strategies from $C$ to local and shared
concerns, respectively, given the current state. $Z$ denotes the
\textit{scope} of a shared concern, \ie the coordinating set (see
\refsec{sec:CoordinationProcedure}).

The objective of coordination is then to find:

\vspace{-4mm}
\begin{equation} \label{eq:CoordFormalObj}
c^\star = \argmin_{c' \in C} \rho(\sigma, c').
\end{equation}

It can be said that $\rho(\cdot)$ \textit{predicts} the effect of
switching the \glspl{aps} of all \glspl{al} to $c'$ and associates to
this effect a cost. Using such predictions to drive adaptation is
common for self-adaptive systems \cite{Calinescu_2015,Baresi_2019}.

Finally, we emphasize that $c^\star = (c^\star_1, c^\star_2, \dots,
c^\star_n) \in C$ is a globally-optimal solution and does not give
guarantees with respect to the optimal satisfaction of local,
\gls{sas}-specific goals. This means that one can encounter cases
where $c^\star_i$ does not optimally satisfy a local concern $O_i$.


The \textit{key idea} of \CoordDCOP is to decompose $\rho(\cdot)$ in a
sum of contributions of \glspl{aps} to local concerns plus a sum of
contributions to shared concerns:

\vspace{-3mm}
\begin{equation} \label{eq:CoordFormalObjDecomposed}
  \rho(\sigma,c') = \sum_{i=1..n} \text{k}_i (o_i,s_i,c_i') \\ +
  \sum_{\substack{k_Z \in K_S, \, o_Z \in O_S \\ s_Z \in S_Z}} \text{k}_Z
  (o_Z,s_Z,c_Z') \,.
\end{equation}

This decomposition is general and is suited for systems with both
loosely-coupled \glspl{aps}, \ie many scopes $Z_i$, but with
cardinality $|Z_i|$ small, and tightly-coupled strategies, \ie few
$Z_i$, but with big $|Z_i|$. Moreover, this decomposition can express
strongly conflicting \glspl{aps} with $k_Z(\cdot) = \infty$, strongly
synergetic strategies with $k_Z(\cdot) = 0$, and any other degree of
compatibility with $k_Z(\cdot) \in (0, \infty)$. The same follows for
local concerns.

\subsubsection{Example} \label{sec:CoordFormExample}

We apply the formalization to the running example. A possible design
of $\rho(\cdot)$ is:



\vspace{-6mm}
\begin{multline} \label{eq:RhoDecomposed}
    \rho(\sigma, c') = k_1 \left(o_1, s_1, c'_1\right) + k_2
    \left(o_2, s_2, c'_2\right) \\ + k_3\left(o_3, s_1, s_2, c'_1,
    c'_2\right) \,,
\end{multline}

where $s_1 \in S_1$, $s_2 \in S_2$ are monitored states; $c'_1 \in
C_1$ and $c'_2 \in C_2$ are \glspl{aps}; $o_1 \in O_1$, $o_2 \in O_2$,
$o_3 \in O_S$ are concerns; $k_1: O_1 \times S_1 \times C_1 \to
\mathbb{R}^+$, $k_1 \in K_1$, $k_2: O_2 \times S_2 \times C_2 \to
\mathbb{R}^+$, $k_2 \in K_2$, $k_3: O_S \times S_1 \times S_2 \times
C_1 \times C_2 \to \mathbb{R}^+$, $k_3 \in K_S$ are models. The
contribution to local concerns is given by $k_1(\cdot)$ and
$k_2(\cdot)$, while the contribution to shared concerns is given by
$k_3(\cdot)$; $k_3(\cdot)$ should ensure that \ACeff and \ICperf or
\ACperf and \ICperf are not simultaneously selected during high
demand, \ie avoid starving \SVone of resources. The monitored states
include the amount and qualities of served videos and ads and the used
resources. The local concerns are with respect to the amount and
qualities of served videos and ads, respectively, while the shared
concern is with respect the overall revenue generated by the web
application.

\subsection{Mapping coordination to DCOP} \label{sec:MappingToDCOP}

\begin{figure}[t]
  \centering
  \newcommand{\drawarchmapecontrol}[2]{
  \draw [mminter] (#1.south) -- (#2.north);
}

\begin{tikzpicture}[
  scale=1.0,
  transform shape,
  node distance=10ex and 7ex,
  font=\rmfamily\footnotesize,
  managedsysarch/.style={
    managedsys,
    rounded corners,
    text width=4.5em,
    align=center,
    node distance=4ex and 2ex,
  },
  managingsysarch/.style={
    managingsys,
    align=center,
    text width=3em,
    node distance=4ex and 4ex,
    font=\rmfamily\small
  },
  labelstyle/.style={
    text width=0.5em,
    font=\footnotesize,
    align=right
  },
  defaultstyle/.style={
    rectangle,
    text width=1em,
    align=center,
    draw
  },
  inputstyle/.style={
    defaultstyle,
    fill=black!15
  },
  analysisstyle/.style={
    defaultstyle,
    text width=1.5em,
    minimum width=2em,
    minimum height=2em,
    fill=white
  },
  outputstyle/.style={
    defaultstyle,
    double
  },
  narystyle/.style={
    circle,
    fill=black,
    minimum width=1ex
  },
  constraint/.style={
    draw,
  },
  every matrix/.style={
    nodes={draw, fill=white, node distance=0ex, minimum height=4ex, anchor=west, font=\footnotesize},
    column 1/.style={text width=1.2em, align=center},
    column 2/.style={text width=1.2em, align=center},
    column 3/.style={text width=1.2em, align=center},
    column 4/.style={text width=1.2em, align=center},
  },
  manager/.style={
    thick,
    fill=black!5,
    draw=black!30
  }
  ]
  \pgfsetlayers{bg,main}
  \node[analysisstyle] (A1) at (0, 0) {$x_{2}$};
  \node[analysisstyle, above=6ex of A1] (A2) {$x_{1}$};
  \draw[constraint] (A1) -- node[midway, right]  (C1) {} (A2);

  \draw[constraint] ($ (A1.south) + (-1.5ex, 0) $) -- ++(0ex, -3ex) -|
  node [midway, right] (C5) {} ($ (A1.south) + (1.5ex, 0) $);

  \draw[constraint] ($ (A2.north) + (-1.5ex, 0) $) -- ++(0ex, 3ex)  -| node [midway, right] (C7) {} ($ (A2.north) + (1.5ex, 0) $);


  \node[right=1ex of A1] (a1) {$a_2$};
  \node[right=1ex of A2] (a2) {$a_1$};

  \begin{pgfonlayer}{bg}

      \node[manager, fit=(A1) (a1)] (Infra) {};
      \node[manager, fit=(A2) (a2)] (App) {};
  \end{pgfonlayer}

  \node[labelstyle, text width=1em, align=right, left=1ex of App] (AppLabel) {\textbf{\SVone}};
  \node[labelstyle, text width=1em, align=right, left=1ex of Infra] (InfraLabel) {\textbf{\SVtwo}};

  \node[below right=-2ex and -6ex of Infra] (Expl1) {$f_{2} (\delta_{Z_2}) = \begin{cases} 15 & \; \text{if } \delta_{Z_2} = \text{\ICeff} \\ 0 & \; \text{if } \delta_{Z_2} = \text{\ICperf} \end{cases}$};
  \node[above right=-2ex and -6ex of App] (Expl2) {$f_{1} (\delta_{Z_1}) = \begin{cases} 10 & \; \text{if } \delta_{Z_1} = \text{\ACeff} \\ 0 & \; \text{if } \delta_{Z_1} = \text{\ACperf} \end{cases}$};
  \node[right=-2ex of C1, anchor=west] (Expl3) {$f_{3} (\delta_{Z_1}, \delta_{Z_2}) = \begin{cases} \infty & \; \text{if } \delta_{Z_1} = \text{\ACeff} \land \delta_{Z_2} = \text{\ICperf} \; \lor \\ & \; \quad \delta_{Z_1} = \text{\ACperf} \land \delta_{Z_2} = \text{\ICperf} \\ 0 & \; \text{otherwise} \end{cases}$};



\end{tikzpicture}
  \caption{A mapping of the running example to \gls{dcop}.}
  \label{fig:DCOPApplicationExample}
  \vspace*{-5mm}
\end{figure}
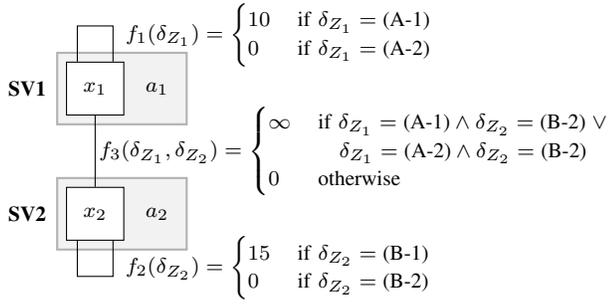

\subsubsection{Overview}

Next, we introduce a mapping of coordination
(\refsec{sec:CoordFormalization}) to a \gls{dcop}. In our approach,
the system integrator specifies, starting from $\langle S, C, O, K,
\Sigma, \rho \rangle$, the structure of a \dcop, including variables,
agents, and constraints. At \runtime, this \dcop is updated
dynamically to take into account the new complete state $\sigma \in
\Sigma$ of the system (see
\refsec{sec:CoordinationProcedure}). Solving the \dcop and obtaining
the optimal coordination solution $c^\star$ is done at \runtime in a
decentralized manner using a distributed constraint optimization
algorithm.




\subsubsection{Variables, domains, and agents}

Each $x_i \in X$ corresponds to the \glspl{aps} selection for
\gls{sas} $i$. The set of possible \glspl{aps} $C_i$ is represented by
the domain $D_i$ of $x_i$. Agents $a_i$ controlling variables $x_i$
and implementing \dcop algorithms are implemented by the \gls{psm}
of each \gls{sas}.


\subsubsection{Mapping concerns to constraints}

It is easy to map \refeq{eq:CoordFormalObjDecomposed} to \dcop
constraints. The contributions to local concerns, \ie $k_i(\cdot)$,
are mapped to \dcop unary constraints, \ie to $f_i$ with scope
cardinality $|S_i| = 1$. We call such constraints \textit{preference
  constraints} because their main purpose is to specify local
preferences towards certain \glspl{aps}. A strong preference for a
certain \gls{aps} $\delta_1 \in D_i$ would be specified as
$f_i(\delta_1) = 0$. If a strategy $\delta_2 \in D_i$ must be avoided,
then $f_i(\delta_2) = \infty$.

We map the contributions of shared concerns, \ie $k_Z(\cdot)$, to
n-ary \dcop constraints, \ie $f_i$ with $|S_i| > 1$. We call them
\textit{consistency constraints} because they quantify \textit{how
  well} two or more strategies work together. For example, if two
\glspl{aps}, specified as $\delta_1 \in D_j \times D_k$ are in strong
conflict, then $f_i(\delta_1) = \infty$. If two strategies, specified
as $\delta_2 \in D_j \times D_k$, are strongly synergetic, then
$f_i(\delta_2) = 0$.

Separating constraints in preference and consistency constraints gives
important privacy benefits. In some algorithms, \eg DPOP
\cite{petcu2005dpop}, agents exchange information about n-ary
constraints with other agents involved in those constraints, while
unary constraints stay private.

\subsubsection{Example}

In the case of high-demand, a possible mapping of coordination to
\gls{dcop} is depicted in \reffig{fig:DCOPApplicationExample}
(variables are white rectangles, agents are gray rectangles, and
constraints are lines). The variable $x_1$ corresponds to the
\gls{aps} assignment of \SVone and has the domain $D_1 = C_1 =
\{\text{\ACeff}, \text{\ACperf}\}$. Variable $x_2$ corresponds to
\SVtwo and its domain is given by $D_2 = C_2 = \{\text{\ICeff},
\text{\ICperf}\}$.

We map $k_1(\cdot)$ and $k_2(\cdot)$ to the preference constraints
$f_1$ and $f_2$, respectively, and $k_3(\cdot)$ to the consistency
constraint $f_3$. The constraint $f_1$ is associated to $x_1$, while
$f_2$ is associated to $x_2$. The consistency constraint $f_3$ high
costs to the cases expected to lead to poor revenue
(\refsec{sec:CoordFormExample}). The solution of this \gls{dcop} would
then be $\delta^*=\left(\text{\ACperf}, \text{\ICeff}\right)$, meaning
that \SVone will acquire more resources during high-demand, while
\SVtwo will adapt the quality of ads.


\subsection{Selecting a DCOP algorithm} \label{sec:SelectAlgorithm}

System designers have a significant number of options when choosing a
\dcop algorithm for coordination. We base the following guidelines on
the theoretical properties of \dcop algorithms and recommendations
from \cite{Fioretto_2018}.

\subsubsection{Constrained resources}

Many \glspl{sas} are constrained by the resources at their disposal,
which affects coordination and the choice of \dcop algorithm. In case
of limited memory, \eg the coordination logic runs on edge devices
\cite{Tsigkanos_2019}, algorithms trading more computation for less
memory usage, \eg the ADOPT search-based algorithm \cite{Modi_2005},
are preferred. In case of intermittent network connectivity, \eg
mobile \glsxtrshort{iot}, it can be better to send fewer messages and
inference-based algorithms like DPOP \cite{petcu2005dpop} are
preferred.


\subsubsection{Complexity of coordination task}

When the behaviors of coordinating \glspl{sas} are tightly-coupled,
\eg robotic systems \cite{Calinescu_2015} with
stringent safety requirements, conflicts between their \gls{aps} can
lead to cycles in the \gls{dcop} constraint graph. Inference-based
algorithms, \eg DPOP, are not suitable for such systems because an
increase in cycles can result in exponential increases in coordination
overheads, \eg size of messages, execution time. Instead, search-based
algorithms are preferred.

\subsubsection{Example}

Because our example does not have any cycles in its \gls{dcop}, the
DPOP algorithm would be a good choice as it has been shown to a very
good performance in such cases.


\newcommand{\valQ}[1]{(\textbf{RQ#1})}
\newcommand{\valQone}{\valQ{1}\xspace}
\newcommand{\valQtwo}{\valQ{2}\xspace}
\newcommand{\valQthree}{\valQ{3}\xspace}

\newcommand{\expOne}{(\textbf{VAREN})\xspace}
\newcommand{\expTwo}{(\textbf{VARPREF})\xspace}
\newcommand{\baselineOne}{\textbf{Baseline 1}\xspace}
\newcommand{\baselineTwo}{\textbf{Baseline 2}\xspace}
\newcommand{\baselineThree}{\textbf{Coordination}\xspace}
\newcommand{\fcoord}{30 days\xspace}
\newcommand{\fpref}{180 days\xspace}

\newcommand{\ami}[1]{$AM_{#1}$}

\newcommand{\istratE}{\text{(I-E)}}
\newcommand{\istratP}{\text{(I-P)}}
\newcommand{\astratS}{\text{(A-S)}}
\newcommand{\astratA}{\text{(A-A)}}
\newcommand{\astratN}{\text{(A-N)}}

\newcommand{\stratperf}{\textit{performance}\xspace}
\newcommand{\stratenerg}{\textit{resource consumption minimization}\xspace}

\newcommand{\estimsimple}{\textit{simple}\xspace}
\newcommand{\estimstats}{\textit{statistics-based}\xspace}
\newcommand{\estimml}{\textit{machine learning-based}\xspace}

\newcommand{\realcost}{\textit{real cost}\xspace}
\newcommand{\expectedcost}{\textit{expected cost}\xspace}

\section{Experimental evaluation} \label{sec:Experimental}

We aim to answer the following research questions:

\begin{enumerate}[left=0pt,label=(\textbf{RQ\arabic*})]
  \item \textit{What is the impact of \CoordDCOP in comparison to
    uncoordinated adaptation?}
  \item \textit{How does \CoordDCOP scale, in terms of communication
    and execution overheads, with increasing system sizes?}
\end{enumerate}

\subsection{Overall experiment design}

\subsubsection{Self-adaptive exemplar} \label{sec:ExpSelfAdaptEx}

To answer \valQone and \valQtwo, we evaluate \CoordDCOP
\footnotemark{} in an exemplar of a \gls{cloud} system consisting of
an infrastructure, providing computational resources, and multiple
applications, consuming the resources. \footnotetext{Code and results of our experiments are available at
  \texttt{\href{https://github.com/pauldragan/acsos2023-dec-coord}{github.com/pauldragan/acsos2023-dec-coord}}.} The \glspl{ms} of the
infrastructure, \aka the \gls{im}, and those of the applications, \aka
the \glspl{am}, coordinate on their \glspl{aps} with respect to costs
and user satisfaction. We decided on this problem setting due to our
interest in \gls{cloud} computing and because similar problems have
been well-studied by the community \cite{Oliveira_2013, Azlan_2015,
  metzger2016coordinated, Nardelli_2019}.



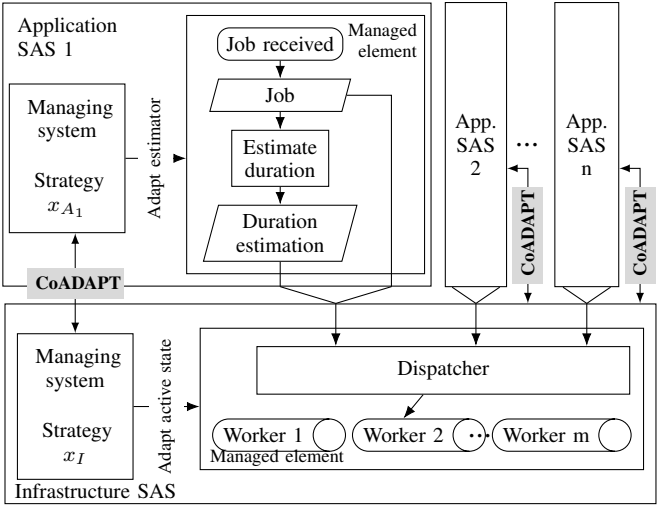
\begin{figure}[t]
  \centering
  \begin{tikzpicture}[
  scale=1.0,
  transform shape,
  font=\rmfamily\footnotesize,
  node distance=3ex and 3ex,
  default/.style={
    draw,
    minimum width=4.5em,
    align=center,
    text width=4.5em,
    text centered,
    on grid
  },
  appman/.style={
    default,
    text depth=10ex,
  },
  estimator/.style={
    draw,
    font=\footnotesize,
    text width=3em,
    align=center
  },
  mape/.style={
    draw,
  },
  cylstyle/.style={
    cylinder,
    draw,
    shape border rotate=90,
    aspect=0.25,
    align=left,
    font=\footnotesize
  },
  worker/.style={
    cylinder,
    draw,
    font=\footnotesize
  },
  job/.style={
    trapezium,
    trapezium left angle=75,
    trapezium right angle=105,
    draw,
    text width=4em,
    align=center,
    minimum height=2ex,
  },
  start/.style={
    rounded corners=1ex,
    draw,
    align=center,
    minimum height=2ex,
  },
  arrowstyle/.style={
    -Latex[]
  },
  coadaptlabel/.style={
    fill=black!15,
    text width=3em,
    align=center,
    font=\scriptsize
  }
  ]


  \node[text width=2.5em, font=\footnotesize, align=center] (AdaptEstimator1) {Strategy $x_{A_1}$};
  \node[text width=3em, align=center, font=\footnotesize, above=1ex of AdaptEstimator1, align=center] (MAPE1) {Managing system};
  \node[draw, inner ysep=0.5ex, fit=(AdaptEstimator1) (MAPE1)] (Managing1) {};

  \node[estimator, right=4em of Managing1] (Estimator1) {Estimate duration};

  \node[job, above=1.5ex of Estimator1] (Job) {Job};
  \node[start, above=1.5ex of Job] (Start) {Job received};
  \node[job, below=1.5ex of Estimator1] (Estimation) {Duration estimation};

  \node[right=2em of Estimator1] (Dummy2)  {};

  \node[draw, inner xsep=2ex, inner ysep=1ex, fit=(Start) (Estimation) (Estimator1) (Job) (Dummy2)] (Managed1) {};
  \node[above right=0.1ex and 0.1ex of Managed1, font=\scriptsize, text width=3em, align=right, anchor=north east] {\Glsxtrlong{me}};
  \node[draw, inner xsep=0.5ex, inner ysep=1ex, fit=(Managed1) (Managing1)] (AM1) {};

  \draw[arrowstyle] (Managing1) to node[midway, font=\scriptsize, rotate=90, fill=white] {Adapt estimator} (\tikztostart -| Managed1.west)  ;

  \draw[arrowstyle] (Start) -- (Job);
  \draw[arrowstyle] (Job) -- (Estimator1);
  \draw[arrowstyle] (Estimator1) -- (Estimation);
  \draw (Estimation) -- (Estimation |- AM1.south) node (Start11) {};
  \draw (Job) -| (Dummy2 |- AM1.south) node (Start12) {};

  \path let
  \p1=(AM1.west),
  \p2=(AM1.east),
  \p3=(AM1.south),
  \p4=(AM1.north)
  in node [
  draw,
  minimum height=\y4-\y3,
  text width=1.65em,
  align=center,
  right=0.5em of AM1,
  ] (AM2) {App. SAS 2};

  \node[font=\large, right=-0.1ex of AM2] (Other) {...};

  \path let
  \p1=(AM1.west),
  \p2=(AM1.east),
  \p3=(AM1.south),
  \p4=(AM1.north)
  in node [
  draw,
  minimum height=\y4-\y3,
  text width=1.75em,
  align=center,
  right=0.5ex of Other,
  ] (AMn) {App. SAS n};


  \path let
  \p4=(AM1.south),
  \p6=($(\p4) + (0, -5ex)$),
  \p7=(Start11.center),
  \p8=($(Start11) + (-2ex, 0ex)$),
  in node [
  draw,
  minimum height=4ex,
  minimum width=14.1em, anchor=north west
  ] (Dispatcher) at (\x8, \y6) {Dispatcher};

  \node[worker, below left=2ex and -2ex of Dispatcher] (Worker1) {Worker 1};
  \node[worker, draw, right=0.5ex of Worker1] (Worker2) {Worker 2};
  \node[worker, draw, right=0.5ex of Worker2] (Worker3) {Worker m};

  \node[draw, inner xsep=1ex, inner ysep=1.5ex, fit=(Dispatcher) (Worker1) (Worker2) (Worker3)] (Managed2) {};

  \draw[arrowstyle] (Dispatcher) -- (Worker2.north);
  \node[font=\large, right=-2.3ex of Worker2] {...};

  \node[text width=3em, align=center, left=5em of Dispatcher, font=\footnotesize] (MAPEIM) {Managing system};
  \node[text width=2.5em, font=\footnotesize, align=center, below=1ex of MAPEIM] (AdaptIM) {Strategy $x_I$};
  \node[draw, inner ysep=0.5ex, fit=(MAPEIM) (AdaptIM)] (ManagingIM) {};

  \node[above left=-0.5ex and -0.5ex of AM1, text width=4em, anchor=north west] {Application SAS 1};

  \node[draw, inner xsep=1ex, inner ysep=2ex, fit=(Managed2) (ManagingIM)] (IaaS) {};

  \node (Meet1) at ( $ (Start11)!0.5!(Start12) $ |- IaaS) {};
  \draw (Start11.center) -- (Meet1 |- IaaS.north);
  \draw (Start12.center) -- (Meet1 |- IaaS.north);
  \draw[arrowstyle] (Meet1 |- IaaS.north) node (SIM1) {} -- (SIM1 |- Dispatcher.north);

  \node (Start21) at ($ (AM2.south) + (-2ex, 0) $) {};
  \node (Start22) at ($ (AM2.south) + (2ex, 0) $) {};
  \node (Meet2) at ( $ (Start21)!0.5!(Start22) $ |- IaaS) {};
  \draw (Start21.center) -- (Meet2 |- IaaS.north);
  \draw (Start22.center) -- (Meet2 |- IaaS.north);
  \draw[arrowstyle] (Meet2 |- IaaS.north) node (SIM1) {} -- (SIM1 |- Dispatcher.north);

  \node (Startn1) at ($ (AMn.south) + (-2ex, 0) $) {};
  \node (Startn2) at ($ (AMn.south) + (2ex, 0) $) {};
  \node (Meetn) at ( $ (Startn1)!0.5!(Startn2) $ |- IaaS) {};
  \draw (Startn1.center) -- (Meetn |- IaaS.north);
  \draw (Startn2.center) -- (Meetn |- IaaS.north);
  \draw[arrowstyle] (Meetn |- IaaS.north) node (SIM1) {} -- (SIM1 |- Dispatcher.north);

  \draw[arrowstyle] (ManagingIM) to node[midway,fill=white, font=\scriptsize, rotate=90, text width=6em, align=center] {Adapt active state} (\tikztostart -| Managed2.west);

  \draw[latex-latex] (ManagingIM) -- node[coadaptlabel,midway] {\textbf{\CoordDCOP}} (ManagingIM.north |- Managing1.south);

  \draw[latex-latex] ($ (AM2.east) + (0, -2ex) $) to ++(1.75ex, 0) to
  node[coadaptlabel,midway,rotate=90]
  {\textbf{\CoordDCOP}} (\tikztostart |- IaaS.north) ;

  \draw[latex-latex] ($ (AMn.east) + (0, -2ex) $) to ++(1.75ex, 0) to
  node[coadaptlabel,midway,rotate=90]
  {\textbf{\CoordDCOP}} (\tikztostart |- IaaS.north) ;


  \node[below left=-2.5ex and -0.1ex of IaaS, anchor=north west] (IaaSLabel) {Infrastructure SAS};
  \node[below left=-2.5ex and -0.1ex of Managed2, anchor=north west, font=\scriptsize] (Managed2Label) {\Glsxtrlong{me}};

\end{tikzpicture}
  \caption{Job-dispatching \gls{sas} exemplar.}
  \label{fig:Exp_System}
  \vspace*{-5mm}
\end{figure}

Concretely, we carry out our experiments in the Simdex exemplar
\cite{Krulis_2022}. Simdex simulates a job-dispatching system where a
backend distributes jobs between workers. Originally, Simdex is a
monolithic \gls{sas} with centralized adaptation logic. To evaluate
\CoordDCOP, we extend Simdex to a setting involving multiple
\glspl{sas}: an infrastructure and $n$ applications
(\reffig{fig:Exp_System}).

The \gls{me} of the infrastructure, consist of workers and a
dispatcher. The dispatcher assigns jobs to workers based on the
workers' availability, load, and a \textit{job duration
  estimation}. The \im adapts workers' state (active or inactive)
according to two possible \glspl{aps}: \stratperf $\istratP$, where
all workers are active at all times, and \stratenerg $\istratE$, where
more workers are activated with increasing demand.


\reffig{fig:Exp_System} includes $n$ applications. The \gls{me} of
each application is an estimator estimating the duration of each
job. The \appm adapts the job duration estimator upon having processed
new jobs to improve future predictions; adaptation is done according
to three possible \glspl{aps}: \estimstats $\astratA$, using a running
average of job durations, \estimml $\astratN$, using a backpropagation
algorithm, and \estimsimple $\astratS$, which does a constant update
using predefined duration limits, regardless of the real duration.

To realize the architecture from \reffig{fig:Exp_System} we reuse the
adaptation logics provided by Simdex, consolidating them as
\glspl{aps}. We extended the simulation to trigger a coordination step
with period $T_R$. To assess adaptation effectiveness, we use the two
metrics provided by Simdex: \textit{resource consumption}, measuring
the number of active workers, and \textit{dissatisfaction}, counting
delayed and late jobs. For more details, see Sections 4.1 and 4.2 in
\cite{Krulis_2022}. Next we apply \CoordDCOP to the subject system.


\subsubsection{Formalizing coordination} \label{sec:SimdexFormalizingCoord}

 First, we formalize coordination by following
 \refsec{sec:CoordFormalization}. We define $\langle S, C, O, K,
 \Sigma, \rho \rangle$:

\begin{itemize}
\item
  $S = S_{I} \times S_{A_1} \times \cdots \times S_{A_n}$, where
$S_{I}$ is the set of worker state and $S_{A_i}$ are the sets of
states of duration estimators,

\item
  $C = C_{I} \times C_{A_1} \times \cdots \times C_{A_n}$, where
$C_{I} = \{\istratE, \istratP\}$ and $C_{A_i} = \{\astratS, \astratA,
\astratN\}$,

\item
  $O = O_L \times O_S$, where $O_L = O_{A_1} \times \cdots \times
  O_{A_n}$, with $O_{A_i} = O_{e_i} \times O_{d_i} \times O_{l_i}$ and
  $O_{e_i}$ are local concerns related to operational costs, $O_{d_i}$
  are related to number of delayed jobs, and $O_{l_i}$ are related to
  number of late jobs; the shared concerns $O_S = O_{1I} \times \cdots
  \times O_{nI}$, with $O_{iI} = O_{e_{iI}} \times O_{d_{iI}} \times
  O_{l_{iI}}$, are between the applications and the infrastructure and
  follow a similar logic with respect to costs, delayed jobs, and late
  jobs; for simplicity, we assume that there are no local concerns at
  the level of the infrastructure,
\item
  $K = K_L \times K_S$, where $K_L = K_{A_1} \times \cdots \times
  K_{A_n}$ with $(e_i, d_i, l_i) \in K_{e_i} \times K_{d_i} \times
  K_{l_i} = K_{A_i}$ giving the contributions of application
  \glspl{aps} to local concerns; and $K_S = K_{1I} \times \cdots
  \times K_{nI}$, with $(e_{iI}, d_{iI}, l_{iI}) \in K_{e_{iI}} \times
  K_{d_{iI}} \times K_{l_{iI}} = K_{iI}$ giving the contributions to
  shared concerns.




\end{itemize}

  Then, $\rho: \Sigma \times C \to \mathbb{R^+} \cup \{\infty\}$ is given by:

\vspace{-5mm}
\begin{multline}
  \rho(\sigma, c') = \sum_{i=1..n} e_i(\sigma_i, c_i') + d_i(\sigma_i,
  c_i') + l_i(\sigma_i, c_i') \\ + \sum_{i=1..n} e_{iI}(\sigma_{iI},
  c_i') + d_{iI}(\sigma_{iI}, c_{iI}') + l_{iI}(\sigma_{iI}, c_{iI}')
  \,,
  \label{eq:RhoSimdex}
\end{multline}

where $\sigma_i \in O_i \times S_i$, $\sigma_{iI} \in O_i \times S_I
\times S_i$, $c_i \in C_i$, $c_{iI} \in C_i \times C_I$.




\subsubsection{Mapping to a DCOP} \label{sec:Exp_Modeling}

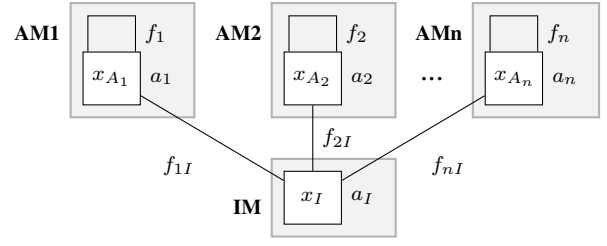
\begin{figure}[t]
  \centering
  \newcommand{\drawarchmapecontrol}[2]{
  \draw [mminter] (#1.south) -- (#2.north);
}

\begin{tikzpicture}[
  scale=1.0,
  transform shape,
  node distance=7ex and 12ex,
  font=\rmfamily\footnotesize,
  managedsysarch/.style={
    managedsys,
    rounded corners,
    text width=4.5em,
    align=center,
    node distance=7ex and 2ex,
  },
  managingsysarch/.style={
    managingsys,
    align=center,
    text width=3em,
    node distance=7ex and 4ex,
    font=\rmfamily\small
  },
  labelstyle/.style={
    text width=2em,
    align=right,
    font=\footnotesize
  },
  defaultstyle/.style={
    rectangle,
    text width=1.2em,
    minimum height=1.2em,
    align=center,
    draw
  },
  inputstyle/.style={
    defaultstyle,
    fill=black!15
  },
  analysisstyle/.style={
    defaultstyle
  },
  planningstyle/.style={
    defaultstyle,
    text width=1.5em,
    minimum width=2em,
    minimum height=2em,
    fill=white
  },
  outputstyle/.style={
    defaultstyle,
    double
  },
  narystyle/.style={
    circle,
    fill=black,
    minimum width=1ex
  },
  constraint/.style={
    draw,
  },
  manager/.style={
    thick,
    fill=black!5,
    draw=black!30,
    inner sep=1ex
  }
  ]

  \pgfsetlayers{bg,main}

  \node[planningstyle, anchor=west] (PAM1) {$x_{A_1}$};

  \node[planningstyle, right=of PAM1] (PAM2) {$x_{A_2}$};

  \node[font=\large, right=2.6em of PAM2] (Other) {...};
  \node[planningstyle, right=of PAM2] (PAM3) {$x_{A_n}$};
  \node[planningstyle, below=5.5ex of PAM2] (PIM) {$x_I$};


  ;

  \draw[constraint] ($ (PAM1.north) + (-2ex, 0) $) -- ++(0ex, 3ex)  -| node [near end, right] (C9) {$f_{1}$} ($ (PAM1.north) + (2ex, 0) $);
  \draw[constraint] ($ (PAM2.north) + (-2ex, 0) $) -- ++(0ex, 3ex)  -| node [near end, right] (C10) {$f_{2}$} ($ (PAM2.north) + (2ex, 0) $);
  \draw[constraint] ($ (PAM3.north) + (-2ex, 0) $) -- ++(0ex, 3ex)  -| node [near end, right] (C11) {$f_{n}$} ($ (PAM3.north) + (2ex, 0) $);

  \draw[constraint] (PAM1) -- node [midway, below left=1ex] {$f_{1 I}$} (PIM);
  \draw[constraint] (PAM2) -- node [midway, right] {$f_{2 I}$} (PIM);
  \draw[constraint] (PAM3) -- node [midway, below right=1ex] {$f_{n I}$} (PIM);





  \node[right=0ex of PAM1] (a1) {$a_1$};
  \node[right=0ex of PAM2] (a2) {$a_2$};
  \node[right=0ex of PAM3] (an) {$a_n$};
  \node[right=0ex of PIM] (aI) {$a_I$};

  \begin{pgfonlayer}{bg}
    \node[manager, fit=(PAM1) (C9) (a1)] (Man1) {};
    \node[manager, fit=(PAM2) (C10) (a2)] (Man2) {};
    \node[manager, fit=(PAM3) (C11) (an)] (Man3) {};
    \node[manager, fit=(PIM) (aI)] (ManIM) {};



  \end{pgfonlayer}

  \node[labelstyle, above left=-4ex and 0ex of Man1] (Label1) {\textbf{AM1}};
  \node[labelstyle, above left=-4ex and 0ex of Man2] (Label2) {\textbf{AM2}};
  \node[labelstyle, above left=-4ex and 0ex of Man3] (Label3) {\textbf{AMn}};
  \node[labelstyle, below left=-4ex and 0ex of ManIM] (LabelIM) {\textbf{IM}};

\end{tikzpicture}
  \caption{\dcop specification of coordination in the exemplar.}
  \label{fig:Experiment_DCOP}
  \vspace*{-5mm}
\end{figure}

We map the coordination formalization to a \gls{dcop} following the
steps from \refsec{sec:MappingToDCOP}. \reffig{fig:Experiment_DCOP}
shows a graphical representation of the resulting
\gls{dcop}. Variables model the \gls{aps} selection for each of the
\glspl{am}, \ie $x_{A_1},\dots,x_{A_n} \in X$, and for the \gls{im},
\ie $x_{I} \in X$. The domains of $x_{A_i}$ are $D_{A_i} = C_{A_i} =
\left\{ \astratS, \astratA, \astratN \right\}$ and the domain of
$x_{I}$ is $D_I = C_{I} \left\{ \istratE, \istratP \right\}$. We map
\refeq{eq:RhoSimdex} to \gls{dcop} constraints as follows. Given the
assignments $\delta_{A_i}$ of variables $x_{A_i}$ and $\delta_{I}$ of
$x_I$, the optimization objective is:

\vspace{-2mm}

\begin{equation}\label{eq:ExpectedCostDecomposed}
    \mu(\delta_I, \delta_{A_1}, \dots) = \sum_{i=1..n} f_{i
      I}(\delta_{A_i},\delta_{I}) + \sum_{i=1..n} f_i (\delta_{A_i})
    \,.
\end{equation}

The consistency constraints between each $x_{A_i}$ and $x_{I}$ are
given by $f_{1 I},\dots,f_{n I} \in F$ with:

\vspace{-4mm}
\begin{equation}\label{eq:CostConsistency}
  f_{i I}(\delta_{A_i}, \delta_I) = e_{iI}(\delta_{A_i}, \delta_I) +
  d_{iI}(\delta_{A_i}, \delta_I) + l_{iI}(\delta_{A_i}, \delta_I) \,.
\end{equation}


\begin{figure*}[t]
  \centering
  \input{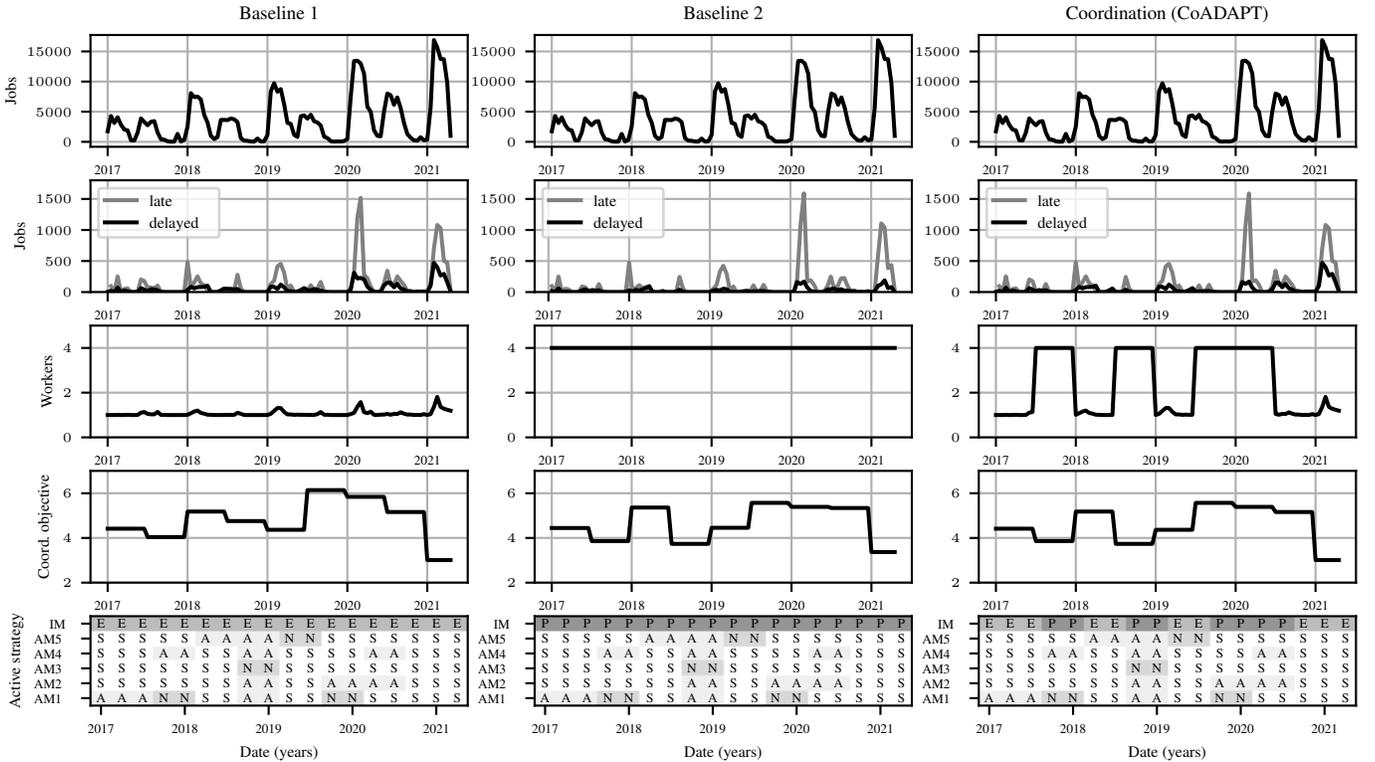}
  \vspace*{-6mm}
  \caption{\textbf{Rows (top to bottom)}: \textbf{Row 1} Workload
    evolution expressed as the number of jobs. \textbf{Row
      2} Number of jobs which have become \textit{delayed} or
    \textit{late}. \textbf{Row 3} Infrastructure utilization expressed
    as the number of active workers. \textbf{Row 4} Evolution of the
    \gls{coordobjective}. \textbf{Row 5} \Glspl{aps} over time, for
    the \gls{im} (E is $\istratE$, P is $\istratP$) and the \glspl{am} (S is
    $\astratS$, A is $\astratA$, N is $\astratN$).  \textbf{(Columns)} Three
    experiments (left to right): \baselineOne, \baselineTwo, and
    \baselineThree.}
  \label{fig:Exp_Pref}
  \vspace*{-5mm}
\end{figure*}

\begin{table}[t]
  \caption{Statistics of the experiments carried out for \valQone. The
    values are averaged over ten runs.}

  \centering
  \begin{tabular}{|l|c|c|c|}
    \hline
    \textbf{Experiment} & \textbf{Jobs delayed} & \textbf{Jobs late} & \textbf{Average workers} \\
    \hline

    Baseline 1 & 1.44\% & 3.61\% & 1.06 \\
    \hline
    Baseline 2 & 0.62\% & 3.03\% & 4.00 \\
    \hline
    Coordination & 1.20\% & 3.44\% & 2.44 \\
    \hline

  \end{tabular}
  \label{tab:Exp_DCOP_results}
  \vspace*{-5mm}
\end{table}

The preference constraints $f_1,\dots,f_n \in F$ are given by:

\vspace{-4mm}
\begin{equation}\label{eq:CostPref}
  f_{i}(\delta_{A_i}) = e_i(\delta_{A_i}) + d_i(\delta_{A_i}) + l_i(\delta_{A_i}) \,.
\end{equation}


We reused some the notation from \refsec{sec:SimdexFormalizingCoord}
for the $e_i$, $d_i$, $l_i$, $e_{iI}$, $d_{iI}$, and $l_{iI}$
functions. As they were not our focus, we realized these functions as
simple key-value maps, where keys are \glspl{aps}; we chose the map
values experimentally to match the effect of strategies assessment
metrics. Important for the experiments is that $e_{iI}$, $d_{iI}$, and
$l_{iI}$ have a linear structure, \eg $e(\delta_{A_i}, \delta_I) =
e_i(\delta_{A_i}) + e_I(\delta_I)$, where $e_I$ gives a contribution
of an infrastructure strategy to energy costs (same logic applies to
$d_{iI}$, and $l_{iI}$).






\subsubsection{Algorithm selection} Because the constraint graph of
the coordination \gls{dcop} is acyclic, we choose the DPOP algorithm
\cite{petcu2005dpop} due to its good performance and privacy
features. We use the implementation of DPOP from the pyDcop library
\cite{rust2019pydcop}.

\subsection{Experiments for answering \valQone} \label{sec:ValQOneSetup}

\subsubsection{Design}

We study the effect of coordination between the \appms and the \im on
the subject system under evolving requirements at the level of the $n$
application \glspl{sas}. Such situations are typical to \gls{sas},
where it is expected that requirements will change during operation
\cite{Carwehl_2023}. To simulate changes in requirements and goals we
introduce small random variations in $e_i$, $d_i$, $l_i$, every
$T_R=180$ days, \ie the coordination period, and update the \gls{dcop}
accordingly.  We choose $T_R = 180$ days so as to reflect that
requirements are slow-changing.

We then compare the values of the assessment metrics
(\refsec{sec:ExpSelfAdaptEx}) when the \appms and the \im coordinate
versus when the \appms and the \im do not coordinate. This results in
three experiments:


\baselineOne. The \glspl{am} change their individual \gls{aps}
assignments in response to new $e_i$, $d_i$, and $l_i$, without
coordinating, \ie they individually optimize their preference
constraints. The strategy of the \im is fixed to \istratE.

\baselineTwo. Same as the previous experiment, but the \gls{aps} of
the \gls{im} is fixed to \istratP.

\baselineThree. The \glspl{sas} coordinate on the \gls{aps}
assignments with \CoordDCOP.



All experiments include \textit{five} applications and \textit{one}
infrastructure. We have decided on these numbers so as to ensure
sufficient complexity, while keeping simulation times to reasonable
durations. The \appms and \im have access to all \glspl{aps} from
\refsec{sec:ExpSelfAdaptEx} and the infrastructure has four workers
(as in the original Simdex paper \cite{Krulis_2022}).


As our experiments involve randomized elements, \ie $e_i$, $d_i$, and
$l_i$, we run all experiments ten times, each run with a different
seed of the random number generator. Then we compute the average of
delayed, late jobs, and number of active workers. Of course, the seed
is same between the \baselineOne, \baselineTwo, and \baselineThree of
the same run.


\subsubsection{Results}


The top row in \reffig{fig:Exp_Pref} shows identical plots depicting
the workload. Note that this plot is only shown for explaining the
peaks in other plots from \reffig{fig:Exp_Pref}. Workload itself does
not drive the assignment of \glspl{aps}; the assignment is done by
minimizing \refeq{eq:ExpectedCostDecomposed}.

Row four of \reffig{fig:Exp_Pref} shows the value of the objective
function change every $T_R = 180$ days due to new
requirements. \baselineOne exhibits different characteristics than
\baselineTwo because of the different contributions of $\istratE$ and
$\istratP$. Interestingly, \baselineThree seems to be a mixture of
\baselineOne and \baselineTwo, \eg in the first half of 2017 the
objective function has the same value as \baselineOne, while in the
second half the value is identical to that of \baselineTwo. Indeed,
the objective function of \baselineThree is always the smallest
between the those of \baselineOne and \baselineTwo; given that
coordination is expressed as a minimization problem, this plot
indicates that coordination behaves correctly.

\begin{figure*}[t]
  \centering
  \input{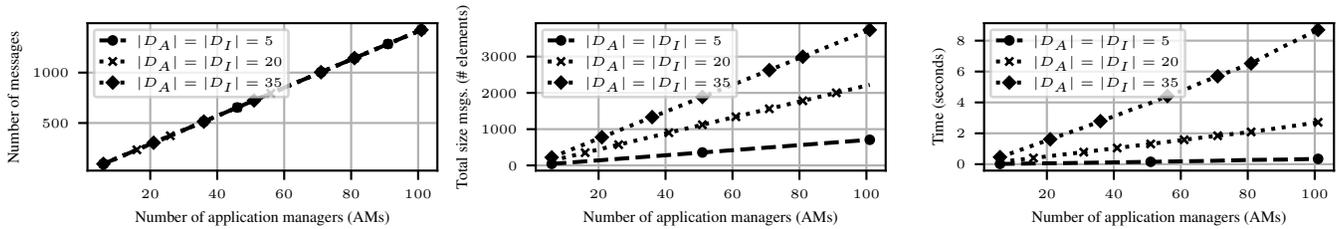}
  \vspace*{-9mm}
  \caption{Effect of increasing the number of \glspl{am} on
    coordination overheads.}
  \label{fig:Exp_Scalability}
  \vspace*{-2.5mm}
\end{figure*}

\begin{figure*}[t]
  \centering
  \input{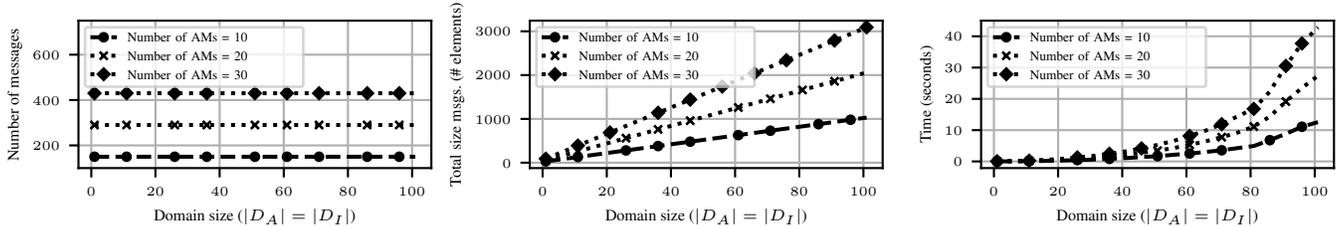}
  \vspace*{-9mm}
  \caption{Effect of increasing the domain sizes on coordination
    overheads. The experiment fixes the number of \glspl{am} to ten.}
  \label{fig:Exp_Scalability_B}
  \vspace*{-7mm}
\end{figure*}

The bottom row of \reffig{fig:Exp_Pref} shows the assigned
\glspl{aps}. Each row corresponds to a specific \appm or to the \im
and cells are the active strategies at a particular point in time. The
strategies selected by the \appms being the same in the three
experiments is explained by the linear construction of $e_{iI}$,
$d_{iI}$, and $l_{iI}$: minimizing the objective function is
equivalent to individually minimizing the terms corresponding to the
\appms\xspace \refeq{eq:CostPref} and then coordinating on the optimal
\im strategy \eqref{eq:CostConsistency}. \baselineThree achieves the
lowest objective function values by always selecting the optimal
\gls{im} \gls{aps} between $\istratE$ and $\istratP$.

The second and third rows of \reffig{fig:Exp_Pref} show the delays in
job processing and average number of active workers, respectively.
Although it is difficult to assess the effects of coordination solely
from these plots, \reftab{tab:Exp_DCOP_results} shows that
\baselineThree achieves better results than \baselineOne but slightly
worse than \baselineTwo in terms of late and delayed jobs. An
explanation is in the third row of \reffig{fig:Exp_Pref}: \baselineOne
attempts to minimize the number of active workers $\istratE$, whereas
\baselineTwo maintains four active workers at all times
$\istratP$. \baselineThree switches between $\istratE$ and $\istratP$,
resulting in periods where all workers are simultaneously active and
periods where workers are active during high demand.
\reftab{tab:Exp_DCOP_results} shows that, indeed, the average number
of active workers in \baselineThree is between \baselineOne and
\baselineTwo.

\subsubsection{Discussion}

These results validate \CoordDCOP: through coordination, the \appms
and \im select their \glspl{aps} so as to jointly minimize the
objective function. One could say, based on \reffig{fig:Exp_Pref} and
\reftab{tab:Exp_DCOP_results}, that the results are not strikingly in
favor of \baselineThree. We should keep in mind two aspects: 1) the
results achieved by \baselineOne and \baselineTwo are already very
good, therefore any improvement will not appear as significant, and 2)
coordination is purely driven by an open-loop objective function,
constructed experimentally. Devising the best objective function for
Simdex was not the focus of this work, but better objective functions
could be considered in the future, \eg by employing performance
modeling or machine learning techniques \cite{Rossi_2020}.

\subsection{Experiments for answering \valQtwo}  \label{sec:ValQTwoSetup}

\subsubsection{Design}

We study the scalability of \CoordDCOP for increasing numbers of
applications and strategies. We carry out this experiment using
\pydcop only, and not Simdex, because we are interested in
coordination overheads and not in the results of coordination. The
experiments run on a system with an Intel i7-1065G7 CPU and 16GB RAM.

\subsubsection{Results}

\reffig{fig:Exp_Scalability} shows the effect of increasing the number
of \appms. For simplicity, we assume $|C_{i}| = |C_{I}|$. To see
the relation between increasing the number of applications and that of
increasing the number of \glspl{aps}, we ran the experiment for three
domain sizes. We observe: 1) the number of exchanged messages
increases linearly with the \appms and is not influenced by the domain
sizes, 2) the total size of messages increases linearly with the
\appms and depends on the domain sizes, \ie larger domains result in
larger messages, and 3) the execution time increases linearly with the
\appms and larger domains give longer execution times.

These results are expected given the characteristics of the DPOP
algorithm and our simple coordination task (acyclic graph -- see
\reffig{fig:Experiment_DCOP}). In DPOP the number of exchanged
messages increases linearly with the number of agents and does not
depend on domain sizes. An increase in the number of \appms gives an
increase in the number of exchanged messages with the same factor. The
total size of messages in DPOP increases exponentially in the worst
case and depends on the number of cycles in the problem graph and on
domain sizes. However, because we have no cycles, we are actually in a
best case scenario, giving a linear increase of total message size. A
similar logic follows for time-complexity.



\reffig{fig:Exp_Scalability_B} shows the effect of increasing numbers
of strategies. The results support previous discussions: the number of
exchanged messages does not depend on the domain sizes and the total
size of messages increases linearly due to our acyclic graph. The
execution time increases quadratically with the domain size. The
reason is that $D_{A_i}$ and $D_I$ are increasing simultaneously,
resulting in a quadratic increase of the time required for evaluating
the consistency constraints $f_{iI}$.


\subsubsection{Discussion}

\CoordDCOP shows promising scalability, with coordination overheads
increasing mostly linearly or quadratically with increasing problem
sizes. \CoordDCOP could effectively coordinate many applications with
an infrastructure for large-scale \gls{cloud} \glspl{sas}.  However,
the DPOP algorithm may limit more complex coordination settings, where
the \dcop constraint graph has many cycles. In such cases, it would be
worthwhile to consider other \dcop algorithms, \eg those exploiting
parallelization \cite{Modi_2005}.






\subsection{Threats to the validity of experiments}

We have identified and addressed the following threats to the
\textit{internal validity}: 1) the size of the validation scenario,
\ie limited number of \appms and \glspl{aps}, which we addressed by
including a scalability experiment, 2) the researchers' bias in
extending Simdex, which we addressed by carrying out minimal changes
to the exemplar and by using the original adaptation logics and
dataset and 3) the choice of coordination objective, which we
addressed by specifying the some goal components as random variables.


The main threat to the \textit{external validity} is the choice of
subject system, which we addressed by focusing on a load-balancing
scenario (highly-relevant to the \gls{sas} community
\cite{Oliveira_2013,Rossi_2020,Nardelli_2019,Chen_Bahsoon2_2017}).
Another threat is the choice of DPOP for \valQone. However, because
DPOP guarantees true optimal solutions, our results will hold for other
algorithms offering similar guarantees, \eg ADOPT.

\vspace{-1mm}



\section{Related work} \label{sec:Related_Work}

\textit{1) Coordination of adaptation planning strategies:}
In contrast to \CoordDCOP, existing techniques
\cite{Wang_Ying_Harmonizing_2007, Khakpour_2010, Glazier_2020} for
coordinating \glspl{aps} are centralized. For example,
\cite{Glazier_2019,Glazier_2020} propose a meta-manager employing a
model checker to find the best assignment of \glspl{aps}. Similar to
us, \cite{Glazier_2019,Glazier_2020} frame coordination as an
optimization problem, however only covering the centralized case and
without differentiating between concerns.

\textit{2) Meta-adaptation}
Assignment of \glspl{aps} is essentially a special case of
meta-adaptation
\cite{Gerostathopoulos_Bures_2017,Lesch_Hadry_2022}. Many
architectures and approaches for meta-adaptation have been proposed,
however, in contrast to \CoordDCOP, they target a single \gls{sas}
\cite{Gerostathopoulos_Bures_2017,Lesch_Hadry_2022} or employ
centralized coordination \cite{Siqueira_2023}.

\textit{3) Decentralized self-adaptive systems:}
A number of coordination techniques were proposed under the term of
\textit{decentralized self-adaptive systems}
\cite{Weyns_2013,Quin_2021}. However, these techniques are usually
conceived from the perspective of a single administrative domain, thus
they do not differentiate between local and shared concerns.
Moreover, existing techniques also suffer from the other limitations
mentioned in \refsec{sec:NewIntro}, \ie they employ a supervisor
\cite{Rossi_2020,Tsigkanos_2019,Gerostathopoulos_2019,Baresi_2019},
share private information
\cite{DAngelo_2020,Azlan_2015,Calinescu_2015}, make unrealistic
assumptions \cite{metzger2016coordinated}, or rely only on local
decision-making
\cite{DAngelo_2020,metzger2016coordinated,Calinescu_2015}.

\textit{4) Collective adaptive systems:}
There exist many coordination approaches for collective adaptive
systems (\eg see
\cite{diaconescu2017architectures,Bucchiarone_2019}). These approaches
focus on self-organization and dynamic coordination of systems that
sporadically, spontaneously, and ephemerally collaborate at
\runtime. These systems coordinate as long as it serves their own
interests, usually lacking a permanent common goal. In contrast,
\CoordDCOP addresses the coordination among individual \glspl{sas}
that cooperate and were assembled together at \designtime to fulfill a
common goal.

\textit{5) DCOP for SAS:}
There are only few works mentioning \dcop for \gls{sas}. In
\cite{harold2019towards} the author suggests that \gls{dcop} could
outperform existing designs for \gls{sas} based on the \gls{mape}
loop. Unlike \cite{harold2019towards}, we propose \gls{dcop} as a
coordination mechanism rather than a replacement for \gls{mape}
loops. Additionally, while we offer detailed specification techniques
and experimental evaluation, \cite{harold2019towards} covers the topic
with limited depth. In their works \cite{Uzonov_2021}, the authors
briefly mention \dcop as a potential component of their architecture
for multi-agent middlewares. Their work does not describe in detail
how \dcop should be employed nor do the authors evaluate their
solution. In terms of application-oriented papers, \cite{Leeuwen_2017}
uses \dcop to realize self-adaptation in the context of wireless
charging and \cite{Hirsch_2021} uses \dcop for self-adaptation in
decentralized manufacturing. In contrast to these works, \CoordDCOP
provides an application-independent coordination technique.

\section{Conclusion and outlook} \label{sec:Conclusion}

We introduced \CoordDCOP, a decentralized technique for coordinating
multiple \glspl{sas}. \CoordDCOP specifies coordination as a
constraint optimization problem and introduces two types of
constraints: 1) preference constraints, expressing concerns local to a
system, \eg adaptation preferences, and 2) consistency constraints,
expressing concerns shared by multiple systems, \eg adaptation
conflicts. The resulting specification, paired with a distributed
optimization algorithm, ensures that local concerns remain private to
a system during coordination, while shared concerns are only known to
other systems involved in those concerns. We have realized \CoordDCOP
for the coordination of \glspl{aps}, demonstrated its feasibility in a
\gls{cloud} computing exemplar, and analyzed experimentally its
scalability.

As future work, we wish to further improve the privacy of \CoordDCOP
and remove the reliance on a system integrator at \designtime. To this
end, we plan to investigate protocols for constructing the constraint
optimization problem in a decentralized way. Moreover, we wish to
evaluate \CoordDCOP in more settings, such as complex,
multi-stakeholder systems that span the whole \gls{cloud} -- IoT
computing continuum.







\bibliography{refs}

\end{document}